\documentclass[twocolumn]{aastex701}

\usepackage{graphicx}
\usepackage{eurosym}
\usepackage{natbib}
\usepackage{booktabs}
\usepackage{amsmath}
\usepackage{graphics,graphicx}
\usepackage{hyperref}
\hypersetup{urlcolor=blue}
\usepackage{cleveref}
\usepackage{eurosym}
\usepackage{booktabs}
\usepackage{wrapfig}
\usepackage{tabularx}
\usepackage{soul}

\newcommand{\rsun}{$R_\odot$}
\newcommand\be{\begin{equation}}
\newcommand\ee{\end{equation}}

\def\PP{{\cal P}}

\def\PP{{\cal P}}

\def\la{\mathrel{\mathchoice {\vcenter{\offinterlineskip\halign{\hfil
$\displaystyle##$\hfil\cr<\cr\sim\cr}}}
{\vcenter{\offinterlineskip\halign{\hfil$\textstyle##$\hfil\cr<\cr\sim\cr}}}
{\vcenter{\offinterlineskip\halign{\hfil$\scriptstyle##$\hfil\cr<\cr\sim\cr}}}
{\vcenter{\offinterlineskip\halign{\hfil$\scriptscriptstyle##$\hfil\cr<\cr\sim\cr}}}}}

\begin{document}

\title{The Shape of the Solar Tachocline}

\author[orcid=0000-0002-6163-3472, sname=Basu, gname=Sarbani]{Sarbani Basu}
\affiliation{Department of Astronomy, Yale University, PO Box 208101, New Haven, CT 06520-8101, USA}
\email[show]{sarbani.basu@yale.edu}
\correspondingauthor{Sarbani Basu}

\author[orcid=0000-0003-1531-1541, sname=Korzennik, gname=Sylvain]{Sylvain G. Korzennik}
\affiliation{Center for Astrophysics $|$ Harvard \& Smithsonian, Cambridge, MA 02138, USA}
\email[show]{skorzennik@cfa.harvard.edu}

\begin{abstract}
 Early helioseismic results have shown that the tachocline has a prolate shape. However, the models used in  {those} studies constrained the tachocline to be either prolate or oblate. We use helioseismic data obtained from long time series (2304 and 4608 days) to determine the shape of the solar tachocline. Like previous work, we use forward modeling methods for this work; however, we allow more flexibility for the shape of the tachocline. {We find that the tachocline does indeed deviate from a simple prolate structure and bulges out at mid latitudes. The center of the tachocline lies in the radiative zone at low latitudes, in the convection zone at intermediate latitudes, and back in the radiative zone at high latitudes. The high-latitude ($ > 60^\circ$) behavior is however, uncertain and model dependent. Models that allow more variation of the shape indicate that the tachocline at high latitudes is almost coincident with the base of the convection zone.}

\end{abstract}


\section{Introduction} 
\label{sec:intro}


One of the early surprises of helioseismic analyses was that solar internal rotation did not look like what was expected. The Sun's rotation rate, it turns out, is not ``constant on cylinders,'' that is with iso-rotation contours parallel to the rotation axis, as early models predicted {\citep[e.g.][]{glatz}}. Instead, the results showed that in the convection zone (CZ) of the Sun differential rotation is nearly constant {as a function of radius}, while the radiative interior rotates like a solid body 
{\citep[see e.g.,][and references therein]{schou}, though it has been shown that between latitudes of 15$^\circ$ and $55^\circ$, the rotation contours make an angle with the rotation axis of about 25 degrees \citep{GilmanHowe}.} Connecting the two zones is a thin shear layer, known as the ``tachocline.''

The origin of the tachocline is not well understood, nor is the role it plays. {However, some models of the solar dynamo cite this layer } as one that is key in the process of magnetic field generation in the Sun and other solar-like stars, since it is a region of strong shear and therefore very capable of converting weak poloidal fields into strong toroidal fields {that the large-scale toroidal magnetic fields generated at the tachocline are responsible for the emergence of sunspots \citep{dikpati1999, chatterjee2004, guerrero2008}}. 
In fact, in 3D MHD simulations of the solar dynamo that include the tachocline, most of the magnetic field develops at the base of the convection zone \citep{guerrero}. It has also been argued that the tachocline plays a key role in establishing the period of the solar cycle, in the origin of torsional oscillations, and the scaling law of stellar magnetic fields as a function of the Rossby number \citep{guerrero2017}. 

Knowledge of the shape of the tachocline may be used to distinguish between different models of the tachocline as well as different solar dynamo theories. For instance, using a purely hydrodynamical model of the tachocline, \citet{balbus}\ claim that the tachocline has a quadrupolar structure (one that goes as $\cos^2\vartheta$, $\vartheta$ being the colatitude).  In many dynamo theories \citep[see e.g.,][]{petrovay}, the shape of the tachocline is important in determining the strength of the magnetic field that can be stored, thus making it a diagnostic for the geometry of the magnetic field. \citet{dikpati2001}, using an MHD version of a shallow-water model, showed that magnetic fields in the tachocline makes it prolate,  i.e., the position of the tachocline at low latitudes is deeper than that at high latitudes. \citet{guerrero2007} found that such a prolate tachocline is able to reproduce solar-like butterfly diagrams.

Early helioseismic investigations on the properties of the tachocline \citep{abc, paulchar} assumed a $\cos^2\vartheta$ dependence of the shape, and hence, the deduced shape was constrained to be either oblate or prolate; both investigations concluded that the tachocline was prolate. There have been efforts at determining the properties of the tachocline at each latitude separately (i.e., by fitting models that are just a function of radius as any given latitude). There the results are mixed; while some results show that the tachocline is prolate \citep{abc}, others \citep{sbhma2000, sbhma2003} seem to show a near discontinuous behavior around a latitude of $30^\circ$. In these studies, the tachocline appears to be at a constant radius at lower latitudes, and at a constant larger radius at higher latitudes. However, the results were not statistically significant; moreover, there are no models that can explain a discontinuous, {or even a near-discontinuous,} tachocline.

Previous work on determining the shape of the tachocline used helioseismic data obtained from relatively short time series. \citet{abc} used data obtained with a 360-day time series, while \citet{paulchar} used data from a 2-year long, but single site, and hence low duty cycle, time series. In this paper, we investigate the shape of the tachocline using solar oscillation data obtained from longer time series than those used in previous work. Fitting long time series reduces the uncertainties in the data, which allows us to put firmer constraints on the shape of the tachocline.

The remainder of this paper is organized as follows. We describe the data used in Section~\ref{sec:data}, our tachocline model and fitting method is described in Section~\ref{sec:model}. We describe our results in Section~\ref{sec:res}, and discuss and summarize our findings in Section~\ref{sec:conc}.

\section{Data used} \label{sec:data}

We use solar oscillation data from the ground-based Global Oscillation Network Group \citep[GONG:][]{gong} along with those obtained by the  Michelson Doppler Imager \citep[MDI:][]{mdi} on board the Solar and Heliospheric Observatory (SOHO) and the Helioseismic and Magnetic Imager \citep[HMI:][]{hmi} on board the Solar Dynamics Observatory (SDO).

We use rotational frequency splittings obtained by an independent data reduction pipeline \citep{syl1, syl2, syl3, syl4, syl5, sgk2018, sgk2023}, which we refer to as the ``SGK'' pipeline.

{The SGK pipeline derives mode parameters from time series of spherical
harmonic coefficients that are multiples of 72 days\footnote{Initially the
GONG pipeline considered fitting time series as short as 36 days, aka the GONG
month. They eventually settled on fitting 108-days long time series, or 3x 36
days, every 36 days, to increase the signal to noise ratio of the resulting
power spectra. The MDI standard pipeline decided to fit 72 day long time series, 
and later the HMI standard pipeline adopted the same convention. The
SGK pipeline adopted the MDI/HMI convention.}.}
For this work, we use splittings obtained with $32\times72$-day (i.e., a bit more than 6 years) time series and with $64\times72$-day (approximately 12.6 years) time series. The GONG, MDI, and HMI projects' pipelines do not produce frequencies and splitting coefficients with such long time series that are needed for this work. {While, in principle, noise can be reduced by averaging the data sets obtained with shorter time series,  we choose not to do this to avoid small differences in the mode sets of the individual sets, and the choices that need to be made about how the splittings from the different constituent sets are weighted. However, there is one HMI set obtained with the official pipeline using a $32\times 72$-day time series that was obtained in the manner described in \citet{tim}, and we use that set as a double-check on our results' independence of the fitting methodology.} For each time series length, we use one set each of MDI and HMI data, and two GONG sets, one covering the same time period as the MDI set and one contemporaneous with the HMI set. The start dates of the $32\times72$-day sets and the $64\times72$-day sets are the same and are listed in Table~\ref{tab:data}.

\begin{table}[]
  \centering
  \caption{Data sets used. The start dates have the \mbox{YYMMDD} format. }
  \scriptsize
\begin{tabular}{lcc}
\toprule
Data Set && Start Date \\
\midrule
GONG Set~1 && 19960501 \\
GONG Set~2 && 20100711 \\
MDI && 19960501 \\
HMI && 20100711 \\
{HMI (Project)\footnote{Only $32\times72$-day data}} && 20100711\\
\toprule
\end{tabular}
\label{tab:data}
\end{table}

{Solar oscillation frequencies are described with three labels: the degree $l$ that defines the number of nodes along the surface, the order $n$, which is the number of nodes in the radial direction, and the azimuthal order $m$ that describes the number of nodes along the equator and can take values from $-l$ to $+l$. In the absence of rotation, or any other feature such as large-scale magnetic fields, all modes with the same $l$ and $n$ have the same frequency, regardless of the value of $m$, making the mode frequencies $2l+1$-fold degenerate. Rotation and magnetic fields lift this degeneracy. The difference in frequency between the $\nu_{ln0}$ and $\nu_{lnm}$ is usually referred to as frequency splittings.} 
The different data sets { that we use} are available in the standard form, i.e.,  with frequencies  expressed as follows:
\begin{equation}
\nu_{nlm}
= \nu_{nl} + \sum_{j=1}^{j_{\rm max}} c_j (n,l) \, \PP_j^{(l)}(m), 
\label{eq:eq_split}
\end{equation}
where $\nu_{nl}$, or the {mean} frequency of a mode of degree $l$ and radial order $n$, is determined by the spherically symmetric part of solar structure, $c_j$ are {\it splitting coefficients}, and $\PP_j$ are re-scaled Clebsch-Gordon coefficients \citep[see][]{ritz}. In this decomposition, the odd-order $c_j$ are caused  by the solar rotation, while the even-order coefficients contain the signature of asphericity and magnetic fields. The different odd-order $c_j$ coefficients have information about the latitudinal distribution of the solar rotation. In order to compare with earlier work, we work with the $c_j$ coefficients as defined by \citet{ritz}, rather than the $a_i$ coefficients that the projects' pipelines use; these coefficients are related and can be transformed from one form to another quite easily {(see \citealt{scdt}, \citealt{pijpers})}. 

The $c_1$ coefficient contains information of the component  of rotation that is independent of latitude; $c_3$ is proportional to  $P_3(\vartheta)$, defined in Eq.~\ref{eq:p5}, where $\vartheta$ is the colatitude, and thus is sensitive to only the prolate or oblate component of solar rotation. Coefficient $c_5$ is proportional to $P_5(\vartheta)$ and hence has a $\cos^4\vartheta$ dependence, while $c_7$ is proportional to $P_7$, which, as can be seen, has a $\cos^6\vartheta$ dependence on latitude.
{Following \citet{ritz},}
\begin{equation}
\begin{split}
P_3(\vartheta) &=5\cos^2\vartheta-1,\\
P_5(\vartheta) &=21\cos^4\vartheta-14\cos^2\vartheta+1, \\
P_7(\vartheta) &=85.8\cos^6\vartheta-99\cos^4\vartheta+27\cos^2\vartheta-1.
\end{split}
\label{eq:p5}
\end{equation}

\section{The tachocline model}
\label{sec:model}

\begin{figure}
    \centering
    \includegraphics[width=3.30 true in]{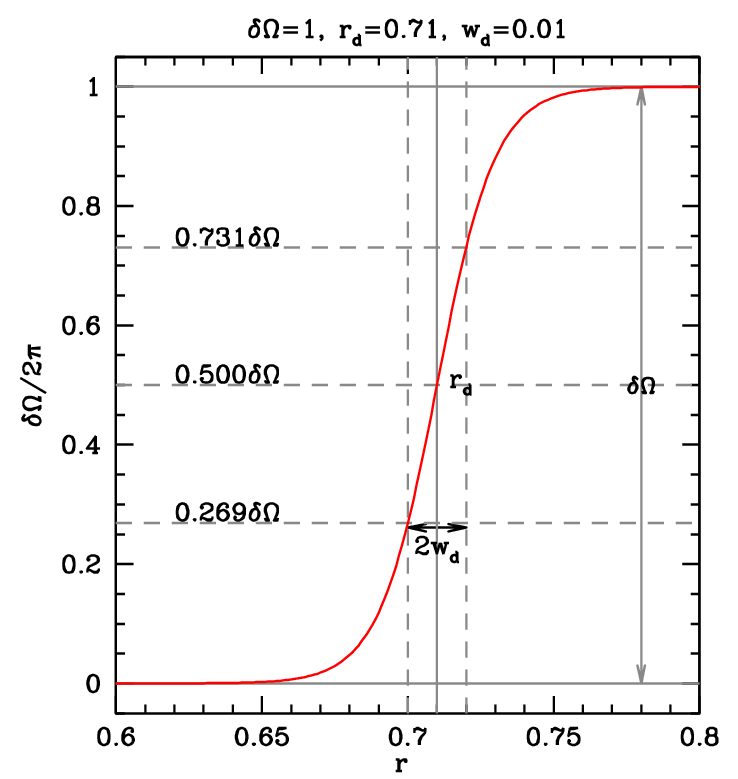}
    \caption{The model of the tachocline. The tachocline parameters used in this figure are marked at the top of the figure.}
    \label{fig:tach}
\end{figure}

For any given latitude, we model the tachocline as a sigmoid following \citet{basu1997}, \citet{abc}, \citet{antiabasu2011}, \citet{basuantia2019} and \citet{basukorz}: 
\begin{equation}
    \Omega(r)_{tach}=\frac{\delta\Omega}{1+\exp[{-(r_d-r)/w_d}]},
    \label{eq:tach}
\end{equation}
where $\delta\Omega$ is the jump in the rotation rate between the convection zone and the interior, $r_d$ is the position of the tachocline, defined as the midpoint of the transition (or discontinuity), { and $w_d$ a measure of the width} of the transition layer. { This parametrization is illustrated in Fig.~\ref{fig:tach}, where one sees that amplitude of the rotation jump changes by 46.2\% over a width of $2\,w_d$ around $r_d$.} For fully 2D fits, the quantities 
$\delta\Omega$, $r_d$ and $w_d$ are modeled as functions of latitude (see Eq.~\ref{eq:2d}).

\begin{figure}
    \centering
    \includegraphics[width=3.25 true in]{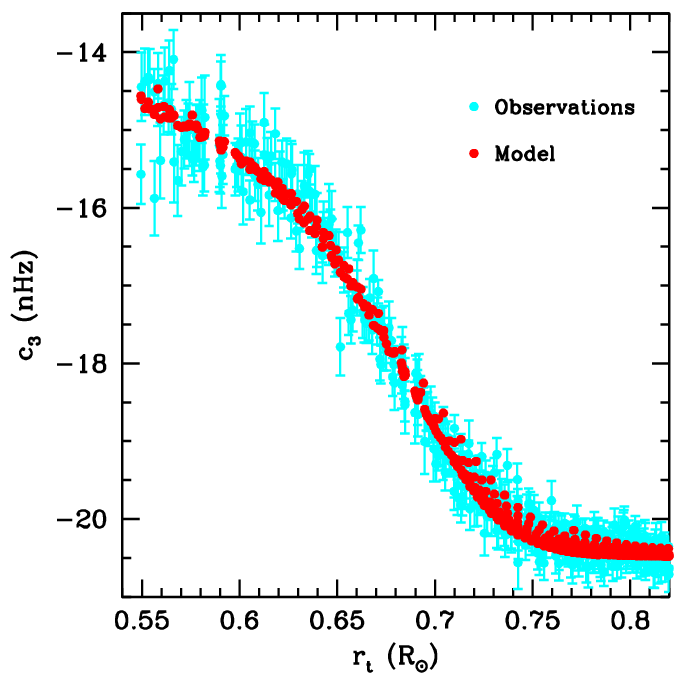}\\
    \includegraphics[width=3.25 true in]{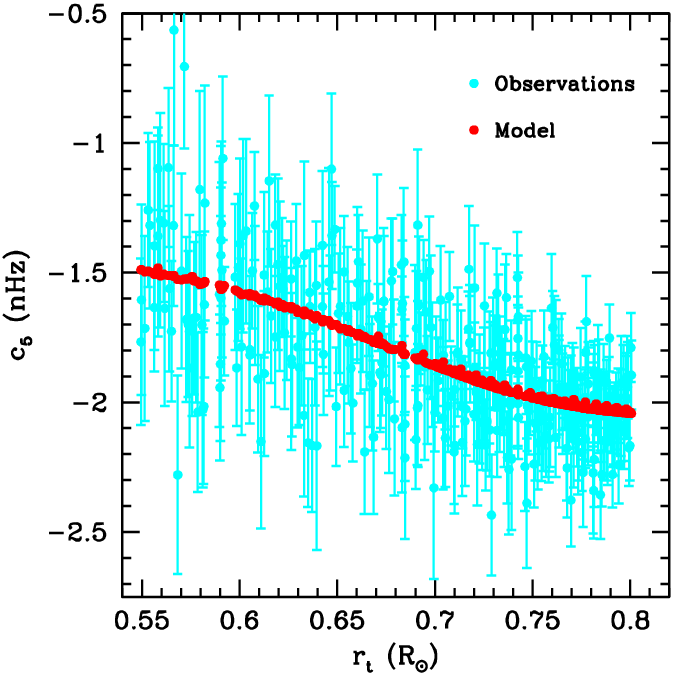}
    \caption{Top: A sigmoid fitted to the $c_3$ splitting coefficient for the HMI $64\times72$-day set. Bottom: a sigmoid fitted to the $c_5$ splitting coefficient of the same set. The background cyan points with error-bars are the observed coefficients plotted as a function of their lower turning point. The red points show the coefficients resulting from our best-fit sigmoid model. The $c_5$ component of the $32\times72$-day sets can also be fitted with a sigmoid. This splitting coefficient is very noisy in shorter sets. }
    \label{fig:c3c5fit}

\end{figure}

In earlier studies (e.g., \citealt{abc}, \citealt{basuantia2019}) the position $r_d$ was modeled as
\begin{equation}
    r_d=r_{d1}+r_{d3}P_3(\vartheta),
    \label{eq:rd}
\end{equation}
which essentially means that they could at most have a $\cos^2\vartheta$ dependence, i.e, they would find that the tachocline was
 {spherically symmetric,} oblate or prolate (since the $c_3$ component has the signature of a tachocline, as was shown by \citealt{agk1996} and \citealt{basu1997} {it is unlikely to be spherically symmetric}). We refer to the above model as the two-term fit to the position of the tachocline.
The latitudinal dependence of $r_d$ was motivated by the fact that the $c_5$ splitting coefficients of the datasets used in the earlier works did not indicate any signature of the tachocline. However, the $c_5$ coefficients obtained from longer time series reveal a clear signature of the tachocline and, as shown in Fig.~\ref{fig:c3c5fit}, they can be fitted with a sigmoid model.
 
Hence we add another term to $r_d$ to model a higher-order latitudinal variation:
\begin{equation}
    r_d=r_{d1}+r_{d3}P_3(\vartheta)+r_{d5}P_5(\vartheta).
    \label{eq:rd5}
\end{equation}
We also fit the tachocline position to the older model, as in Eq.~\ref{eq:rd}. We refer to this as the three-term model of the position $r_d$.

The full 2D model that we use is the same as that used by \citet{basuantia2019}, i.e., 
 \begin{align}
 \Omega(r, \vartheta)=\begin{cases}
 {\Omega_c+ \Omega_{\rm tach}} \\ \quad\quad\quad\quad\mbox{if } r\le0.7R_\odot\\
\Omega_c +B(r-0.7)
+\Omega_{\rm tach}\\
\quad\quad\quad\quad\mbox{if }  0.7 < r\le0.95R_\odot\\
\Omega_c+0.25B -C(r-0.95)
+\Omega_{\rm tach}\\
\quad\quad\quad\quad\mbox{if } r>0.95R_\odot, 
\end{cases}
\label{eq:2d}
\end{align}
where  $\Omega_c$, $B$ and $C$ are free parameters and $\delta\Omega_{tach}$ is given by Eq.~\ref{eq:tach}. 
The position, 
$r_d$, is given either by Eq.~\ref{eq:rd} or Eq.~\ref{eq:rd5} (i.e., a two-term or three-term latitudinal expansion), while $w_d$ is modeled as $w_{d1}+w_{d3}P_3$, and
$\delta\Omega$ as $\Omega_3 P_3 +\Omega_5 P_5$. 
We tried to fit an $\Omega_7 P_7$ term to our sets, but the term was unconstrained except for the results from the fit to the HMI $64\times72$-day set.

We use simulated annealing  \citep{anneal1,anneal2}  {to obtain}  a {minimum $\chi^2$}  fit between the observed and computed values of the splitting coefficients $c_1$, $c_3$ and $c_5$. This algorithm uses randomly generated values of the fitting parameters. We assume that { the random values}  have Gaussian  {distributions}, with their mean and width determined from existing inversions of rotational splittings. These inversions have clearly shown the presence of a tachocline, but do not fully resolve it, since in general regularization is achieved via some form of smoothing.  Given that there is a chance that the solution becomes trapped in a local minimum, we make 100 different realizations using different sequences of randomly selected initial guesses in order to derive a global $\chi^2$ minimum. We can be certain that the algorithm reached a global minimum by inspecting the likelihood-weighted distribution of all the parameters for all iterations, where we defined the  likelihood as  $\exp(-\chi^2/2)$. This distribution is single-peaked when a global $\chi^2$ minimum is reached; otherwise, it will have multiple peaks, or be flat if the parameter cannot be constrained. The uncertainties are determined using the traditional bootstrapping method of simulating many realizations of the observations, fit each one of them in exactly the same manner as the original data and use the spread as a measure of uncertainty \citep{Laarhoven+Aarts}. 

We only use a subset of the modes for our work, namely the ones most sensitive to the tachocline. We restrict ourselves to using  modes with frequencies between 1.5 mHz and 3.5 mHz that have lower turning points\footnote{
The lower turning point is the deepest location a mode penetrates, according to ray theory, and effectively the depth where the oscillation sensitivity decreases drastically; mode properties are most affected by the structure and dynamics at that depth.}
between 0.55\rsun\ and 0.85\rsun\ for the range of degrees that is covered by the data set. 
This subset gives good coverage of the tachocline while keeping the  uncertainties low; this also removes the need to properly account for the near-surface shear layer in the tachocline model, although, as Eq.~\ref{eq:2d} shows, we do use a crude model of that layer.

\section{Results}
\label{sec:res}

\subsection{The shape of the tachocline}
\label{subsec:shape}

\begin{figure*}
    \centering
    \includegraphics[width=0.45\linewidth]{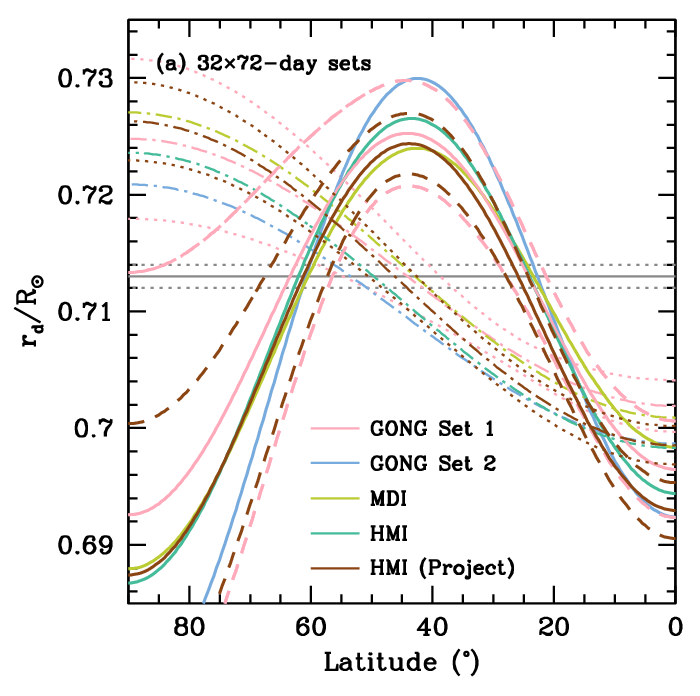}\includegraphics[width=0.45\linewidth]{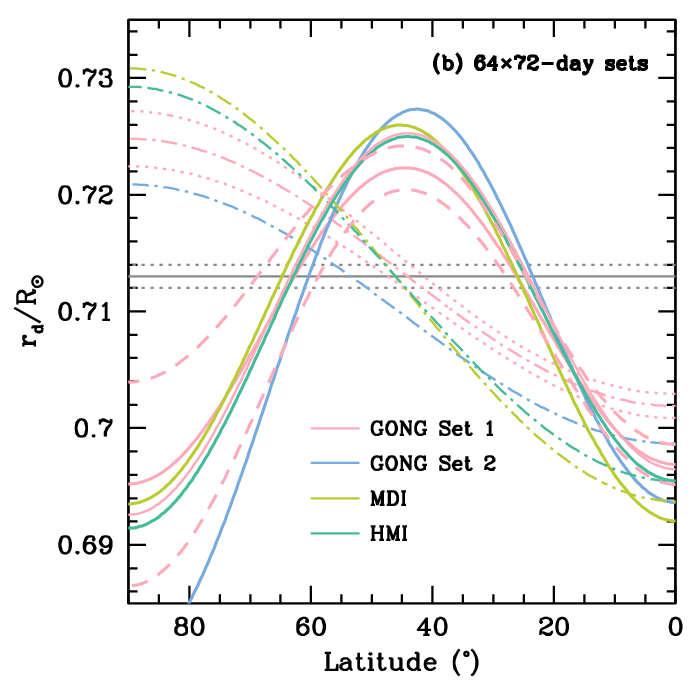}
    \caption{The position of the tachocline plotted as a function of latitude for all the data sets. The solid gray horizontal line shows the position of the convection-zone base, with  the gray dotted lines showing the 1$\sigma$ uncertainty.  Panel (a) shows the results for the $32\times72$-day sets, while panel (b) shows the results for the $64\times72$-day sets. In each panel the solid lines show the results for the three-term case, while the dot-dashed lines are the results when using the two-term latitudinal model. Note how they agree at low latitudes. We show the 1$\sigma$ error limit only for the GONG~Set~1, and the HMI~(Project) results for the sake of clarity. These are marked as dotted lines for the two-term case and dashed lines for the three-term one. The uncertainties are similar for the three other data sets  { obtained with the SGK pipeline}.  {It should be noted that the results close to the pole are essentially an extrapolation, since the data do not have sensitivity there. Also note that the results from the HMI project's pipeline and the SGK pipeline agree well within $1\sigma$ uncertainties.}}
    \label{fig:rad32_64}
\end{figure*}

We fitted the 2D tachocline model with both the two- and three-term latitudinal expansion of the position to the $32\times72$-day set, as well as to the $64\times72$-day set. The results are shown in Fig.~\ref{fig:rad32_64}. Including a third term in the latitudinal expansion of the position of the tachocline makes a statistically significant improvement to the quality of the fit, especially for the $64\times72$-day sets.

One can draw a number of conclusions from Fig.~\ref{fig:rad32_64}. The first, which has been seen before, is that the tachocline does not coincide with the base of the convection zone. It lies below the convection-zone base at low latitudes. Unlike the convection zone which has a very small asphericity,  {deviations of $\la 0.0001 R_\odot$ from the average position} \citep{asph}, the tachocline has a significant   {asphericity}. We also see from the longer data sets, that the tachocline is not merely prolate in shape. It bulges into the convection zone at mid latitudes, and dips into the radiative zone at higher latitudes, although the latter is only a 1$\sigma$ result. 

\subsection{The extent of the tachocline}
\label{subsec:extent}

\begin{figure}
    \centering
    \includegraphics[width=3.25 true in]{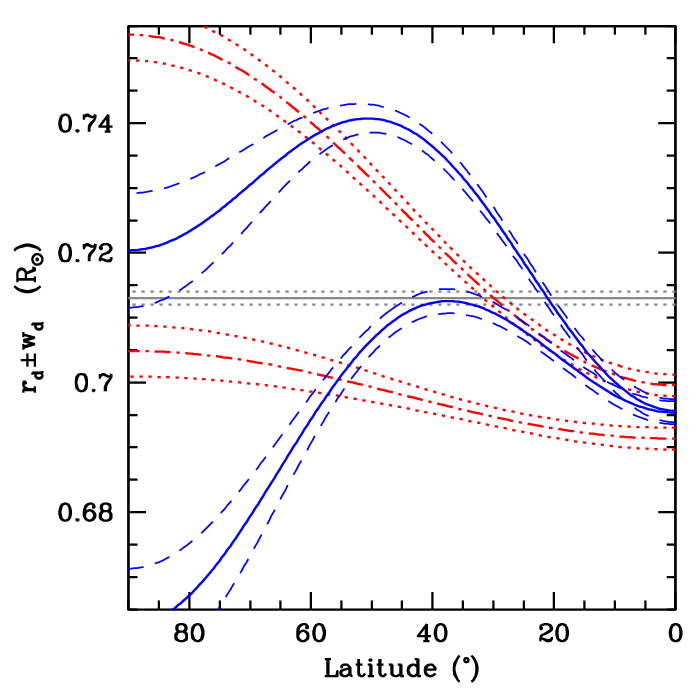}
    \caption{The extent of the tachocline, defined as the region between $r_d-w_d$ and $r_d+w_d$ as obtained with the HMI $64\times72$-day set. The region between  the red dot-dashed lines shows the result of the two-term case, with the dotted lines indicating $1\sigma$ uncertainties. The region between the blue solid lines shows the results of the three-term case, with the dashed lines showing $1\sigma$ uncertainties.}
    \label{fig:extent}
\end{figure}

It is clear from the figure that the tachocline becomes much thicker at higher latitudes. If we take the extent into account, the difference between the two-term and three-term fits becomes much smaller, well within $1\sigma$, indicating that earlier results that only assumed a prolate tachocline are not completely invalid.

\subsection{A comparison with 1D fits}
\label{subsec:1d}

When the position of the tachocline is determined from results of inversions to determine the internal rotation profile,  it is usual to fit a 1D model of the tachocline to latitudinal cuts of the inferred rotation rate. We can do something similar to model the tachocline properties at different latitudes by taking the appropriate combinations of the splittings and fitting those to a 1D model. We do so for the HMI $64\times72$-day set in order to determine whether the three-term case is indeed a better representation of the solar tachocline.  We use the model of \citet{abc}\ to model the tachocline at any given latitude:
\begin{align}
 \Omega(r)=\begin{cases}
 {\Omega_c+ B(r-0.7)+\Omega_{\rm tach}} \\ \quad\quad\quad\quad\mbox{if } r\le0.95R_\odot\\
\Omega_c+0.25B -C(r-0.95)
+\Omega_{\rm tach}\\
\quad\quad\quad\quad\mbox{if } r>0.95R_\odot, 
\end{cases}
\label{eq:1d}
\end{align}
where $\Omega_c$, $B$, and $C$ are three free parameters, in addition to the three parameters, $r_d$, $w_d$ and $\delta\Omega$, used to define $\Omega_{\rm tach}$
(Eq.~\ref{eq:tach}).

\begin{figure}
    \centering
    \includegraphics[width=3.25 true in]{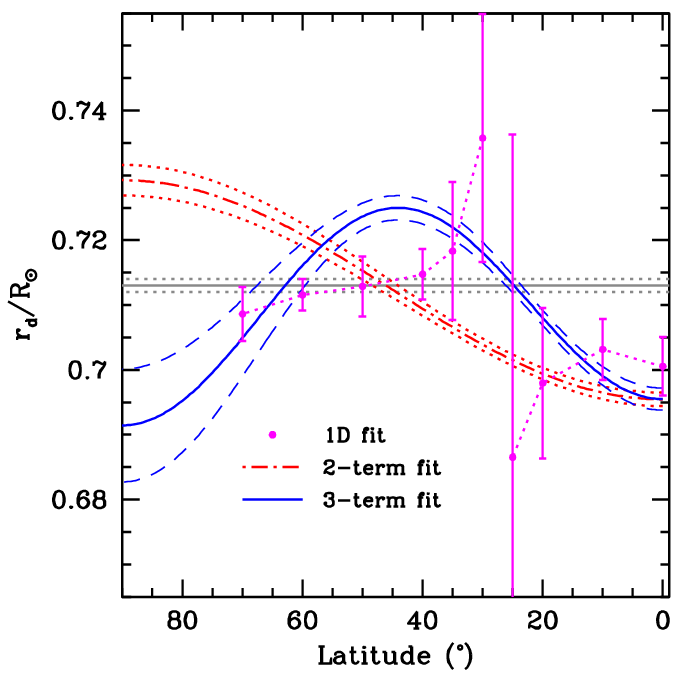}
    \caption{The position of the tachocline obtained by fitting the 1D model shown in Eq.~\ref{eq:1d}\ to the HMI $64\times 72$-day set is plotted as points with 1$\sigma$ error bars;   { the points below and above $25^\circ$} have been connected separately with a dotted line to guide the eye.  The red dot-dashed line is the result of the two-term 2D fit; $1\sigma$ uncertainties are shown as red dotted line. The blue solid line is the result of the three-term 2D fit, with the blue dashed line marking $1\sigma$ uncertainties.}
    \label{fig:1d}
\end{figure}

We show the results of the 1D fit to the tachocline at a few different latitudes, along with the results obtained with the two- and three-term cases in Fig.~\ref{fig:1d}. The 1D results are marginally more consistent with the three-term case, than the two-term one. However, the 1D results appear to show what appears to be  a discontinuity at a latitude of between $25^\circ$ and $35^\circ$ where the jump, $\delta\Omega$ across the tachocline becomes very small. This makes the fits difficult to constrain, resulting in large error bars. 

\begin{figure}[htb]
    \centering
    \includegraphics[width=3.25 true in]{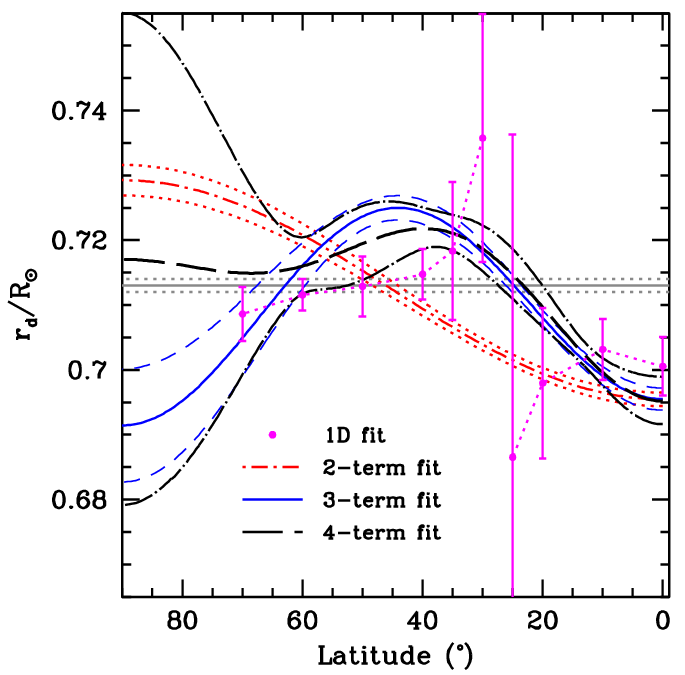}
    \caption{As in Fig.~\ref{fig:1d} we show the position of the tachocline obtained by fitting the 1D model as points with error bars; the fit is to the HMI $64\times72$-day set. The red dot-dashed line is the result of the two-term 2D fit; $1\sigma$ uncertainties are shown as the red dotted line. The blue solid line is the result of the three-term 2D fit, with the blue dashed line marking $1\sigma$ uncertainties. The new, four-term 2D fit is shown with large black dashes, with the large-dash-dotted black lines marking the $1\sigma$ uncertainties.}
    \label{fig:p7}
\end{figure}

The 2D functions that we have used cannot fit a discontinuity; however, we can fit a form that could reveal smaller-scale variations, and hence we fit the model in Eq.~\ref{eq:2d}\ using a four-term latitudinal expansion for $r_d$:
\begin{equation}
    r_d=r_{d1}+r_{d3}P_3(\vartheta)+r_{d5}P_5(\vartheta)+r_{d7}P_7(\vartheta).
    \label{eq:rd7}
\end{equation}
For this four-term model, we need to fit the $c_7$ splitting coefficients in addition to $c_1$--$c_5$.

The results and the comparison with the 1D fits, are shown in Fig.~\ref{fig:p7}. As can be seen, this fit does  better than the others, with the match with the 1D results improving in the intermediate latitudes, and they are within 1$\sigma$. However, even with data from such a long time series (i.e., $64\times72$ days), the uncertainty in the result becomes very large at high latitudes; however, this increase in uncertainty is consistent with what is seen in inversions for the solar rotation profile.  The quality of the fit to the data is similar to  that for the three-term expansion, with very little change in the best-fit $\chi^2$ per degree-of-freedom,  {a change from 1.25 to 1.20,} despite the extra free parameter. Thus it appears that the shape of the tachocline is not as simple as had been believed based on older results.

\section{Discussion and Conclusions}
\label{sec:conc}

We have used helioseismic data obtained with long time series to determine the position of the tachocline, which we define as the midpoint of the region where the rotation rate changes from the convection-zone value to the value in the radiative interior. 

 \citet{abc} and \citet{paulchar}, who had done the first analyses, had modeled the position of the tachocline assuming a $\cos^2$ dependence on the colatitude. Specifically, \citet{abc} modeled  the position as given in Eq.~\ref{eq:rd}, while  \citet{paulchar} modeled the position as $r_d(\vartheta)=r_{d,0}+r_{d,1}\cos^2\vartheta$. As is clear, this constrains the tachocline to be prolate, oblate, or spherically symmetric (if $r_{d,1}=0$). We relaxed the oblateness/prolateness constraint to examine higher-order terms. We found that the data are  fit better with a higher-order, $\cos^4\vartheta$, term added to the model of the tachocline. The sign of the term makes the tachocline bulge out into the convection zone at intermediate latitudes. Note that this term is quite unconstrained if we try to fit data obtained from shorter time series. Modeling efforts suggest that the equilibrium shape of the tachocline has a mid-latitude bulge when there is a very strong magnetic field ($>200$ kG) at low latitudes  {(M. Dikpati \& P. A. Gilman, submitted)}. Thus, our results suggest that a strong magnetic field is present at low latitudes. 

 Our attempts to fit an even higher-order term ($\cos^6\vartheta$) were not completely successful; while the results are good at low latitudes, the uncertainties at high latitudes become extremely large. The quality of the fit is no better than that of the three-term model.

 We also tried to fit 1D models at different latitudes. The uncertainty on the results is large, particularly around the latitude where the radial shear vanishes before changing sign. The results are more consistent with our four-term 2D model (i.e., the one with the $\cos^6\vartheta$ dependence) than with the others. The inconsistency between the 1D and 2D models could just be due to the fact that the tachocline is forced to be very smooth in the 2D models that we have used. The 1D results are similar to the 1D results of \citet{sbhma2000} and \citet{sbhma2003}, who have speculated that the position of the tachocline could be discontinuous, with a constant value below a latitude of $30^\circ$ and a different one above. Although there is no theoretical justification of a discontinuous tachocline, we  fitted a 2D model with the position and width defined as in \citet{sbhma2000}:
 \begin{align}
 r_d=\begin{cases}
 r_0   &\mbox{if } \vartheta \ge \vartheta_0\\
 r_0+r_1 &\mbox{if } \vartheta < \vartheta_0,
\end{cases}
\label{eq:rdisc}
\end{align}
and
\begin{align}
 w_d=\begin{cases}
 w_0   &\mbox{if } \vartheta \ge \vartheta_0\\
 w_0+w_1 &\mbox{if } \vartheta < \vartheta_0,\\
\end{cases}
\label{eq:wdisc}
\end{align}
where $r_0$, $r_1$, $w_0$ and $w_1$ are free parameters, $\vartheta_0$ is the colatitude at which $\delta\Omega$, the jump in the tachocline is 0. The rest of the terms were the same as those in Eq.~\ref{eq:2d}. Guided by our 2D fits, we assumed $\vartheta_0=65^\circ$, and used the $64\times72$-day HMI data. We found that the quality of the fits was not much better and that the $\chi^2$ of the fits was somewhat larger than that for our three-term (i.e., prolate) case, despite having the same number of free parameters. \citet{sbhma2000}\ and \citet{sbhma2003} had found marginally lower $\chi^2$ values in the discontinuous case for GONG data, but not for MDI, and hence these results are consistent in behavior with what had been found before. We get $r_0=0.699\pm0.002R_\odot$ and $r_1=0.022\pm 0.003R_\odot$, thus the results are consistent with all others at low latitudes. However,  the higher latitude value $r_d=0.721\pm0.004$ is inconsistent with the 1D result at the $1\sigma$ level, even though the 1D results were the motivation for this particular exercise.  { Note that varying the colatitude at which the discontinuity occurs in the range $55$--$65^\circ$, does not change the conclusions,} The result
is more consistent with the results of the four-term case. This indicates that the apparent discontinuity in the tachocline position that the 1D fits show is most likely to be a result of not constraining the properties properly at latitudes where the jump in the rotation rate is small.

\begin{figure}[t]
    \centering
    \includegraphics[width=3.35 true in]{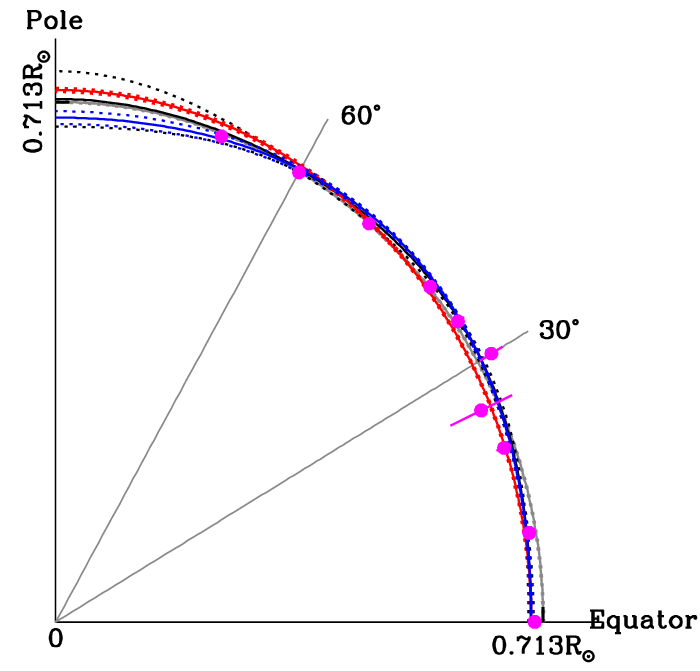}
    \caption{Plot showing the position of the tachocline obtained with the HMI $64\times72$-day data set. The red curve is the result from the two-term fit, the blue curve is that from the three-term fit, and the black is from the four-term fit. Dotted lines of the corresponding colors show 1$\sigma$ uncertainties.  The points with error bars are results of the 1D fits. The position of the convection-zone base is shown in gray.}
    \label{fig:section}
\end{figure}

Since the line plots of $r_d$ as a function of radius do not give an intuitive view of what the center of the tachocline looks like and exaggerate the differences between the models, we show the results obtained with the HMI $64\times72$-day set in Fig.~\ref{fig:section} as a quadrant plot. Note that the base of the convection zone, marked in gray, is spherically symmetric on this scale and forms a reference for the shapes. All results are consistent at low latitudes ($\la 25^\circ$), and show that the center of the tachocline lies in the radiative zone. The models with the best goodness-of-fit criteria indicate that between latitudes of about $25^\circ$ and $60^\circ$ the tachocline lies within the convection zone. The results are more ambiguous at higher latitudes; the 1D, and three-term cases indicate that the center of the tachocline resides in the radiative zone, while the four-term case shows that the tachocline essentially lies at the convection-zone base. 

Our investigation leads us to conclude that the tachocline is not prolate in shape and that it has a more complicated shape than a simple $\cos^2$ colatitude would imply, namely that the tachocline bulges out into the convection zone at mid latitudes. While the shape at high latitudes remains uncertain, the results at low and intermediate latitudes should constrain models of interfacial dynamos.

\clearpage

\begin{acknowledgments}
 { We thank Tim Larson for providing us the $32\times72$-day HMI data set}.
This work is supported by NASA grant 80NSSC23K0563 to SB.
and NASA grants 80NSSC22K0516 and  NNH18ZDA001N-DRIVE to SGK.
This work utilizes GONG data obtained by the NSO Integrated Synoptic Program, managed by the National Solar Observatory, which is operated by the Association of Universities for Research in Astronomy (AURA), Inc., under a cooperative agreement with the National Science Foundation and with contribution from the National Oceanic and Atmospheric Administration (NOAA). The GONG network of instruments is hosted by the Big Bear Solar Observatory, High Altitude Observatory, Learmonth Solar Observatory, Udaipur Solar Observatory, Instituto de Astrofísica de Canarias, and Cerro Tololo Interamerican Observatory. This work also uses data provided by the SOHO/MDI consortium. SOHO is a project of international cooperation between ESA and NASA. We also use data from the Helioseismic and Magnetic Imager on board the Solar Dynamics Observatory. HMI data are courtesy of NASA/SDO and the  HMI science team.

\end{acknowledgments}

\facilities{GONG, MDI, HMI}




\end{document}